# Deep learning-based quality filtering of mechanically exfoliated 2D crystals

Yu Saito[1,*,†,#], Kento Shin[2,†], Kei Terayama[1,3], Shaan Desai[1], Masaru Onga[4], Yuji Nakagawa[4], Yuki M. Itahashi[4], Yoshihiro Iwasa[4,5], Makoto Yamada[1,6,7], Koji Tsuda[1,2,8,*]

[1] *RIKEN Center for Advanced Intelligence Project (AIP), Tokyo 103-0027, Japan,*

[2] *Department of Computational Biology and Medical Sciences, The University of Tokyo, Kashiwa 277-8561, Japan*

[3] *Graduate School of Medicine, Kyoto University, Kyoto 606-8501, Japan*

[4] *Quantum-Phase Electronics Center (QPEC) and Department of Applied Physics, The University of Tokyo, Tokyo 113-8656, Japan*

[5] *RIKEN Center for Emergent Matter Science (CEMS), Wako 351-0198, Japan*

[6] *Department of Intelligence Science and Technology, Kyoto University, Kyoto 606-8501, Japan*

[7] *PRESTO, Japan Science and Technology Agency (JST), 4-1-8, Honcho, Kawaguchi-shi, Saitama 332-0012 Japan*

[8] *National Institute for Materials Science (NIMS), Center for Materials research by Information Integration, Tsukuba, 305-0047, Japan*

*Corresponding author: yusaito@ucsb.edu (Y.S.), tsuda@k.u-tokyo.ac.jp (K. Tsuda.)

†These authors equally contributed to this work.

# Present Address: California NanoSystems Institute, University of California, Santa Barbara, Santa Barbara, CA 93106 USA

**ABSTRACT**

Two-dimensional (2D) crystals are attracting growing interest in various research fields such as engineering, physics, chemistry, pharmacy and biology owing to their low dimensionality and dramatic change of properties compared to the bulk counterparts. Among the various techniques used to manufacture 2D crystals, mechanical exfoliation has been essential to practical applications and fundamental research. However, mechanically exfoliated crystals on substrates contain relatively thick flakes that must be found and removed manually, limiting high-throughput manufacturing of atomic 2D crystals and van der Waals heterostructures. Here we present a deep learning-based method to segment and identify the thickness of atomic layer flakes from optical microscopy images. Through carefully designing a neural network based on U-Net, we found that our neural network based on U-net trained only with the data based on 24 images successfully distinguish monolayer and bilayer $MoS_2$ with a success rate of 70%, which is a practical value in the first screening process for choosing monolayer and bilayer flakes of $MoS_2$ of all flakes on substrates without human eye. The remarkable results highlight the possibility that a large fraction of manual laboratory work can be replaced by AI-based systems, boosting productivity.

Two-dimensional (2D) crystals[1], such as graphene, transition metal dichalcogenides, and van der Waals heterostructures[2] have been intensively studied in a wide range of research fields since they show significant properties that has not been observed in their bulk counter parts. Examples include materials engineering such as electronics and optoelectronics devices[3–7], solid state physics including superconductors[8] and magnets[9,10], chemistry[11,12] and biomedical applications[13]. To manufacture such 2D crystals with atomic layer thickness, especially monolayer or a few-layers samples, mechanical exfoliation, chemical exfoliation, chemical vapor deposition and molecular beam epitaxy have been introduced[11]. Among them, mechanical exfoliation has been instrumental to 2D materials research because it enables us to obtain highly crystalline and atomic 2D layers as is exemplified by ultraclean and high mobility devices based on exfoliated 2D materials combined with the encapsulation by $h$-BN. Recently, high-throughput identification of various unexplored 2D materials via machine learning and the development of a machine for mechanically exfoliated 2D atomic crystals to autonomously build van der Waals superlattices[14] have been reported. These advancements are suggestive of a new research direction for 2D materials that is aimed to explore efficiently enormous materials' properties in large scale using robotics and machine learning. In such a situation, a rapid and versatile method for layer number identification of mechanically exfoliated atomic-layer crystals on substrates is highly desirable in their fundamental research and practical applications.

However, the bottleneck of using mechanical exfoliation is that, through the process, not only do we produce desirable atomic layers (mostly monolayer or bilayer) but also many impractical thicker flakes making it tough to quickly separate the useful layers from the unwanted. To identify the thickness of 2D crystals, several methods such as atomic force microscopy (AFM)[15], Raman spectroscopy[16,17], and optical microscopy (OM) [18,19] are used. AFM is one of the most versatile methods to measure the thickness of various 2D materials, but it takes a relatively long time to measure one region. In addition, the measured value strongly depends on the offset conditions[15] due to, for example, bubbles beneath the sample. OM is nowadays a widely utilized technique to measure the thickness of 2D crystals based on the optical contrast between atomic layers and the substrate[20]. In fact, we can determine the thickness of 2D atomic layers using the contrast difference between the flakes and substrate obtained from the brightness profile of color or grayscale images[18]. The methods explained above, however, need manual work and take a relatively long time to identify atomic layer thickness, and are therefore inappropriate for studying various kinds of materials.

Recently emerged deep learning, a machine learning technique via deep neural networks, has shown immense potential for regression and classification tasks in a variety of research fields[20–25]. In particular, deep neural networks have been very successful in image recognition tasks such as distinguishing images of cats and dogs with high accuracy, and also several physical problems in theoretical physics, for example, detection of phase transitions[26,27] and searching for exotic particles in high-energy physics[28]. Given this, deep neural networks provide an alternative pathway to quickly identifying layer thicknesses of 2D crystals from OM images.

Here, we introduce a versatile technique to autonomously segment and identify the thickness of 2D crystals via a deep neural network. Using a deep neural network architecture, we reproduced the images of 2D crystals from the augmented data based on 24 OM images of $MoS_2$ used as training data and found that both the cross validation score and accuracy rate for the test images through deep neural networks is surprisingly over 70**%**, which means that our model can distinguish monolayer and bilayer with a practical success rate for initial screening process. The present results suggest the deep neural network can become another promising way to quickly identify the thickness of various 2D crystals in an autonomous way, suggesting a large fraction of manual laboratory work can be dramatic decreased by replacing AI-based systems.

Figure 1 shows an overview of data collection using the OM and the training architecture of the deep neural network. First, we prepared the OM images of $MoS_2$ flakes. Bulk single crystals of $MoS_2$ crystals (SPI Supplies, USA) were employed for the preparation of mechanically exfoliated 2D atomic layers, which were then transferred onto the 300-nm-thick $SiO_2$/ Si substrates in the air. The bright-field OM (BX 51, Olympus) was used to locate and take pictures (optical microscope images) of the $MoS_2$ thin flakes on the Si/$SiO_2$ substrate with one hundred magnification. Each OM image includes different number of flakes with various thicknesses under the different light intensity conditions, we then confirmed the thickness and layer number of each samples using AFM and contrast based on OM images, and prepared 35 OM images, which were randomly divided into the training data set (24 images) and test images (11 images), the latter of which were prepared to compare prediction with a non-expert human as discussed below, by dividing into three regions of monolayer (blue), bilayer (green) and others (black). It is noted that as we mention below, we used 960 images which are augmented from 24 images as a training data set.

Next, we constructed one of the deep neural network architectures, U-Net, which is based on the fully convolutional encoder-decoder network[29]. The encoder extracts the small feature map from the input image by convolution and pooling layers, and decoder expands it to the original image size by convolution and up-sampling layers. Figure 2 shows an overview of our network. The encoder and decoder are composed of 14 and 18 layers respectively. The encoder consists of 4 repeated layers set which consist of 3×3 convolutions, each followed by a rectified linear unit (ReLU) activation and 2×2 max pooling with stride 2 for downsampling. At each downsampling step, the number of the feature map is doubled. The decoder consists of 4 repeated layers set which consist of 2×2 upsampling convolution and 3×3 convolution followed by ReLU. We added 50% dropout layers after each of the first three upsampling convolution layers. At the final layer of the decoder, a 1×1 convolution converts the feature map to the desired number of classes and softmax activation is then applied. To transmit high-resolution information in the input images, the network has skip connections between each convolutional layer of the encoder and the corresponding upsampling layer of the decoder. Each skip connection simply concatenates all channels at the layer of the encoder with one of the decoders. The dimension, width × height × channels, of the input image is 512×512×3 and it changes to 256×256×64, 128×128×128, 64×64×256, 32×32×512, and 16×16×1024 at each downsampling step in the encoder respectively, and changes in reverse order through the upsampling steps in the decoder.

In the present study, we used the augmented data based on 24 original OM images and corresponding segmentation maps to train the network and use 11 images to evaluate the performance. Here, we particularly choose the classification of $MoS_2$ monolayer and bilayer because the number of training data set is limited to expand the classifications to thicker multilayer. We employed the data argumentation technique, which is one of learning techniques widely used for deep neural networks to improve learning accuracy and prevent overfitting [20,30–33]. By augmentation with randomly cropping, flipping, rotating and changing the color of the original images[34], we increased the training data up to 960 data points. Indeed, this kind of random manipulation and augmentation of the original data is useful for averaging the difference of contrast and number of flakes in each original image. For the color changing augmentation, we converted images to grayscale or adjusted the contrast of the image. With this augmentation, we increased the data to 960 points. Finally, we normalized all pixels between 0 and 1 as preprocessing before training. We then used cross entropy as a loss function of the multi-class classification and also used a weighted cross entropy in order to balance the frequency of each class. The normal cross entropy $E$ is computed as

$$E = \sum_{i=0}^{N}\sum_{k=0}^{K} y_{ik} \log(x_i) \ .$$

Here, $x_i$ is the output of the softmax activation of the U-Net at pixel $i$, $y_i$ and $y_{ik}$ are the one hot vector of the label at pixel $i$ and $k$-th element (scalar quantity) of $y_i$, respectively, such that only the element at the position of the true label is 1 and the others are 0, K is the number of labels, and $N$ is the number of pixels in the image. Similarly, weighted cross entropy $E_w$ is computed as

$$E_w = \sum_{i=0}^{N}\sum_{k=0}^{K} w_k y_{ik} \log(x_i) ,$$

where $w_k$ is the weight of the class $k$. To compute the weight $w_k$, we use median frequency balancing[35], where the weight assigned to a class is the ratio of the median of class frequencies computed on the entire training set divided by the class frequency. This implies that smaller classes in the training data get heavier weights. In the training, we employed minibatch SGD[36,37] and Adam solver[38]. The learning rate, the batch size and the number of epochs were set to 0.00001, 1 and 100, respectively[29]. Also, we confirmed that learning rate, the batch size and the number of sizes is optimized by changing each parameter in the rage of 0.00001 ~ 0.001, 1 ~ 100 and 10~500, respectively.

Figure 3 shows four examples of the original OM images, the segmented images and generated images using U-Net with weighted loss. The blue, green and black regimes show monolayer, bilayer and other parts including thicker flakes and substrate, respectively. It seems that the training neural network can pick up the information of color contrast (against back ground color of substrate) as well as the edge thickness/sharpness, which is indeed useful information for the real screening process. We calculated receiver operating characteristic (ROC) and precision-recall (precision-true positive rate) curves of two-class classifications for the monolayer/others (blue) and bilayer/others (red) identification and confirmed the learning performance in Fig. 4a and b. The ROC curve shows sharply below the false positive rate of 0.1 and then saturates to 1.0 and precision-recall curve show the high precision value (> 90%) below recalls of 0.76 and 0.94 for monolayer and bilayer, respectively, both of which show the high performance of the U-Net with weighted loss.

To perform further evaluations, we investigated and compared the difference of the cross-validation score and accuracy rate for the test images by changing the learning process using the grayscale images and/or contrast adjusted images. To evaluate the performance of the trained U-Net, we used both cross-validation score and an accuracy score on the test images. In the cross-validation, we performed three-fold cross validation, in which a mean of the pixel-wise accuracy in each class was used as an evaluation metric. On the other hand, in the test images, we prepared the 11 images (test images), each of which has problem region to be answered. To check whether the U-Net can predict the test image correctly or not, we predict the class of each pixel of the image by using U-Net and define the class whose pixel is most in the problem region as the predicted class of that region, then compare it with the true class.

We summarize the cross-validation scores for each learning process in Table 1. The U-Net with weighted loss using contrast adjusted augmentation shows the highest score of all, 0.789, which is much higher than that of a normal U-Net without any options. The value of 0.789 is surprisingly high, considering that while a deep neural network needs thousand or even more training points, our case used the data based on 24 training data points. This remarkable result allows us to generate a practical number of training sets which can initially be prepared by lab works using OM. This result also suggests that once we prepare training data sets and perform leaning for a 2D material, we can obtain a tool that can quickly identify atomic layer thickness, monolayer/bilayer/other thicker parts, with an accuracy rate of almost 80 %, which is practically value that can be helpful for the initial screening process. More importantly, in practice, it is not necessary to check and distinguish of all monolayer or bilayer candidates on the substrate, but just needed to pick up some amount of monolayer or bilayer with a high accuracy. According to the precision-recall curve in Fig. 3d, at miss rates of 24% and 6.0% (1-true positive rate), which means that the U-net miss the flakes with a target layer number (monolayer or bilayer), it can distinguish monolayer and bilayer with an accuracy of 90%. This value is practically high for the real experiments. These high success rates of the identifications mean that the present technique based on the U-net potentially can apply other 2D materials on various wafers because it is very rigid against external conditions (e.g., the number of flakes surrounding the target flakes and the light intensity of optical microscope.), and can detect sensitively the color contrast of the surface against back ground color and edge thickness from the optical microscope images.

Finally, to compare the accuracy rate between U-Net and non-expert humans, we calculated the accuracy rate using 11 randomly selected test images (see Supplementary Materials). We found that the accuracy score of U-Net with weighted loss model is 0.733 in all cases, which is higher than the that of a normal U-Net model. Such a tendency is observed in cross validation score, which indicates that the U-Net with weighted loss is better than the normal U-Net for both segmentation and layer number identification tasks. We then compared the accuracy score with that of twelve researchers who were not familiar with 2D materials (non-expert human). The researchers were given three minutes to learn the training data set (see Supplementary Materials for details) and were then asked to respond with the layer numbers (mono- and bilayer) for each segment of new images, which were the same as the test images used in determining the accuracy score for the deep neural network. The accuracy score for humans was $0.67 \pm 0.11$, which was comparable value to that of a U-Net with weighted loss. The present results suggest that the deep neural network based on U-Net with weighted loss is a new tool to rapidly and autonomously segment and identify number of layers of 2D crystals with an accuracy rate comparable to non-expert humans, indicating it can be an essential tool to significantly decrease manual work in the laboratory by boosting the first screening process, which has been usually done by human eyes.

In conclusion, we introduce a versatile technique to autonomously segment and identify the thickness of 2D crystals via a deep neural network. Constructing an architecture consisting of convolutions, U-Net, we reproduced the images of 2D crystals from the 24 OM images of $MoS_2$ and found that both the cross-validation score and the accuracy rate of generated data through U-Net is over 70 percent, which is comparable with non-expert human level. This means that our neural network can distinguish monolayer, bilayer and other thicker flakes of $MoS_2$ on $Si/SiO_2$ substrates with the practical accuracy in the first screening process for searching desirable before further transport/optical experiments. The present study highlights that deep neural networks have great potential to become a new tool for quickly and autonomously segmenting and identifying atomic layer thickness of various 2D crystals and opening a new way for AI-based quick exploration for manufacturing 2D materials and van der Waals heterostructures in large scale.

**Data availability**
The data that support the findings of this study are available from the corresponding author upon reasonable request.

**Acknowledgements**

This work was supported by the 'Materials research by Information Integration' Initiative (MI2I) project and Core Research for Evolutional Science and Technology (CREST) (JSPS KAKENHI Grant Numbers JPMJCR1502 and JPMJCR17J2) from Japan Science and Technology Agency (JST). It was also supported by Grant-in-Aid for Scientific Research on Innovative Areas 'Nano Informatics' (JSPS KAKENHI Grant Number JP25106005) and Grant-in-Aid for Specially Promoted Research (JSPS KAKENHI Grant Number JP25000003) from JSPS. M.O. and Y.M.I. were supported by Advanced Leading Graduate Course for Photon Science (ALPS). Y.N. was supported by Materials Education program for the future leaders in Research, Industry, and Technology (MERIT). M.O. and Y.N. were supported by the Japan Society for the Promotion of Science (JSPS) through a research fellowship for young scientists (Grant-in-Aid for JSPS Research Fellow, JSPS KAKENHI Grant Numbers JP17J09152 and JP17J08941, respectively). M.Y. was supported by JST PRESTO (Precursory Research for Embryonic Science and Technology) program JPMJPR165A

**Author contributions**

Y.S., K. Terayama, M.Y. and K. Tsuda conceived the idea, designed and supervised the Y.S., K.S., K. Terayama implemented the proposed method and analyzed the experimental results. Y.S., M.O., Y.N and Y.M.I. exfoliated $MoS_2$ and collected images of atomic layers. All authors discussed the results. Y.S., K.S., K. Terayama and K. Tsuda wrote the manuscript with contributions from all other co-authors.

**Notes**

The authors declare that they have no competing interests.

**Figure Captions**

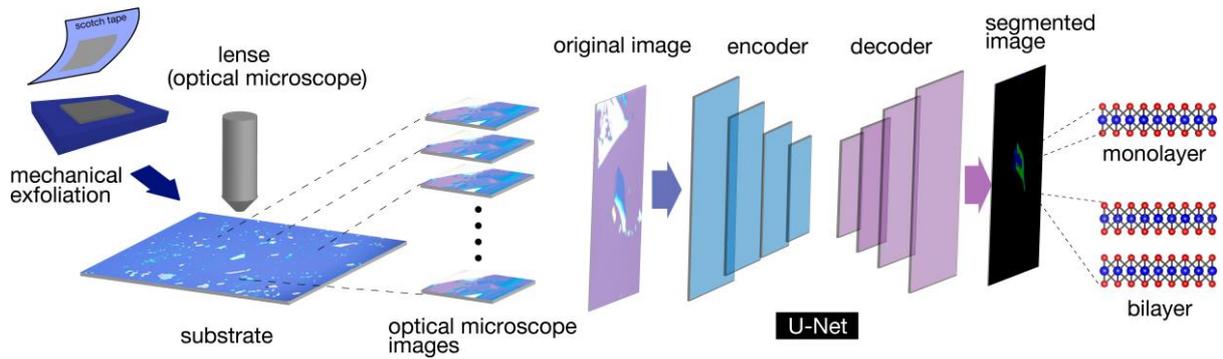

**Fig. 1. Process of quality-filtering of 2D crystals via deep neural network.** First, we mechanically exfoliate $MoS_2$ crystals on a $SiO_2/Si$ substrate and take images using optical microscope. After recording data, we train deep neural networks to generate images from optical microscope images using segmented images. Here, our targets are monolayer and bilayer crystals. For training, we prepared segmented data, which is divided into monolayer (blue), bilayer (green) and other parts (black).

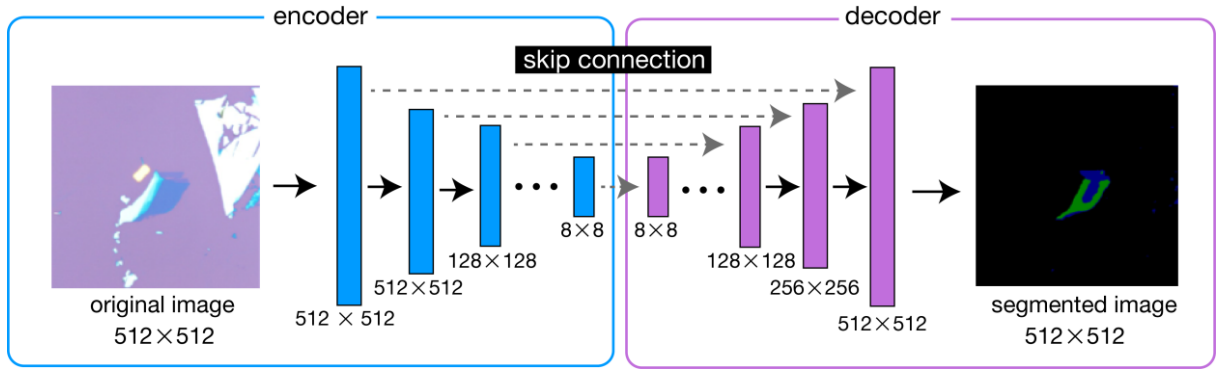

**Fig. 2. Network architectures based on U-Net.** The encoder of U-Net extracts the small feature map from the input image by convolution and pooling layers, and decoder expands it to the original image size by convolution and up-sampling layers. Skip connections are added between each layer of the encoder and the corresponding layer of the decoder in order to transmit a high-resolution information to the decoder. In the architecture, the encoder and decoder are composed of 14 and 18 layers respectively. The dimension, width × height × channels, of the input image is 512×512×3 and it changes to 256×256×64, 128×128×128, 64×64×256, 32×32×512, and 16×16×1024 at each downsampling step in the encoder respectively, and changes in reverse order through the upsampling steps in the decoder.

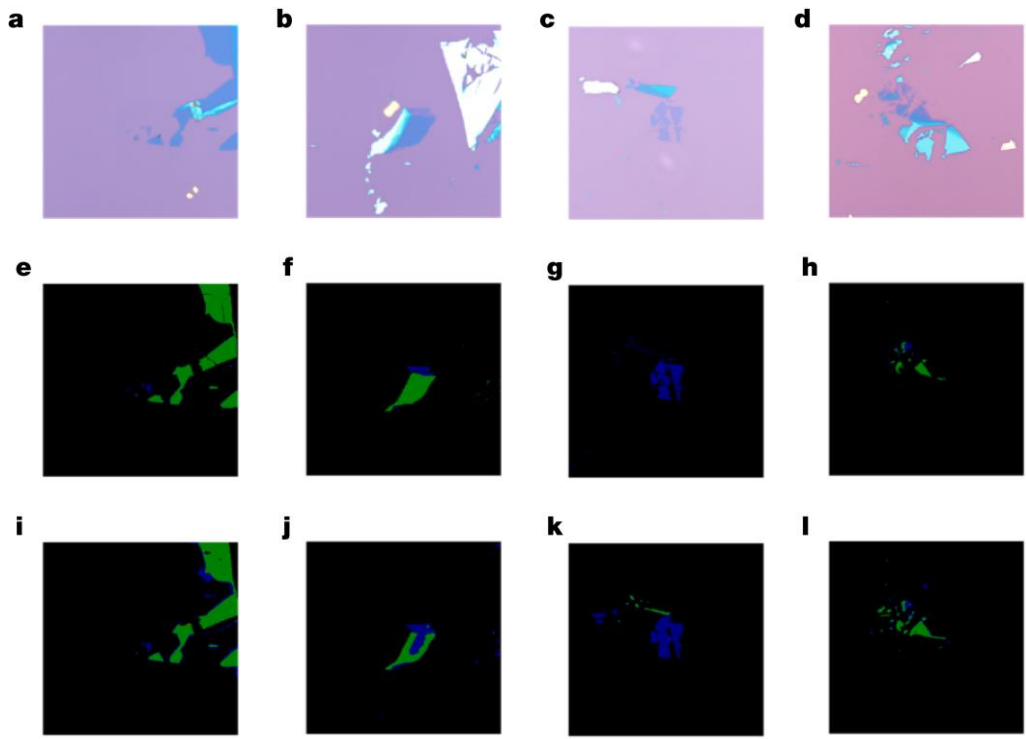

**Fig. 3 Examples of original optical microscope images (a-d), segmented images (e-h) and generated images based on U-Net with weighted loss (i-j).** In the segmented and generated images, blue and green region show monolayer and bilayer region, respectively.

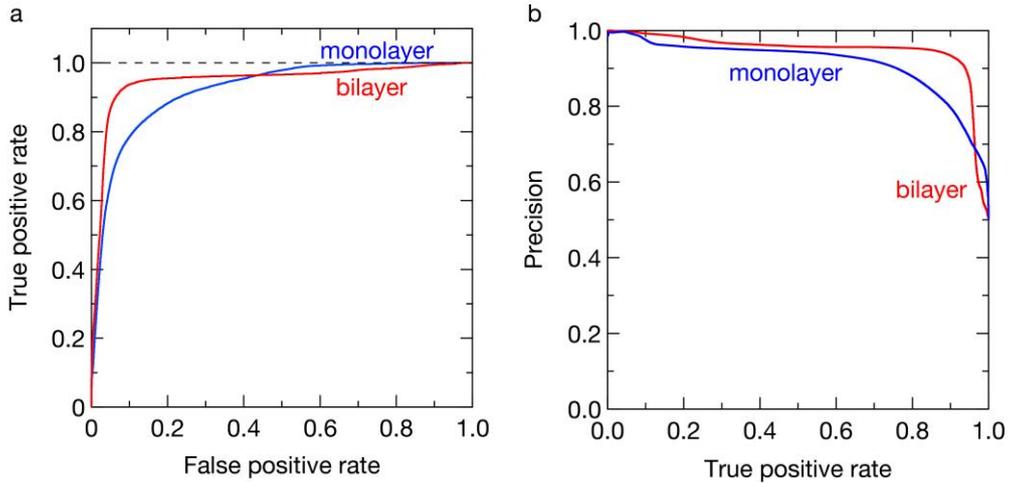

**Fig.4. Receiver operating characteristic (ROC) curves and precision-recall curves for the monolayer and bilayer identification. a,** ROC curve for the monolayer and bilayer identification. Blue and red curve show the ROC for two-class classifications of monolayer/others and bilayer/others, respectively. This ROC curve shows the high performance of the U-Net with weighted loss because the ROC curve rises sharply below the false positive rate of 0.1 and then saturates to 1.0. **b**, Precision-recall (true positive rate) curves for the monolayer and bilayer identification.

**Table 1. Cross-validation scores for U-Net and U-Net with weighted loss.**

|  | U-Net | U-Net with weighted loss |
|---|---|---|
| w/o grayscale & w/o contrast adjusted images | 0.584 | 0.643 |
| w/ grayscale | 0.46 | 0.56 |
| w/ contrast adjusted images | 0.678 | 0.789 |
| w/ grayscale & contrast adjusted images | 0.633 | 0.723 |